\documentclass[showpacs, amsmath, reprint, nofootinbib, showkeys, jcp, aip, longbibliography]{revtex4-2}
\usepackage{dcolumn}    
\usepackage{bm}         
\usepackage{hyperref}
\hypersetup{%
    colorlinks=true, 
    linkcolor=blue,
    urlcolor=blue,
    citecolor=blue, 
}
\usepackage{amssymb,lineno,amsfonts}
\usepackage{graphicx}   
\usepackage{bbm}
\usepackage{cuted}
\usepackage{epstopdf}
\usepackage{setspace}
\usepackage{float}
\usepackage{natbib}
\usepackage{subfigure}
\usepackage{array}
\usepackage{multirow}
\usepackage{booktabs}

\newcommand{\ket}[1]{\left\vert{#1}\right\rangle}
\newcommand{\bra}[1]{\left\langle{#1}\right\vert}

\newcommand{\inpr}[2]{\left\langle{#1}\middle\vert{#2}\right\rangle}

\newcommand{\norm}[1]{\left\vert{#1}\right\vert^2}
\newcommand{\tr}{\mathrm{Tr}}

\begin{document}

\title{Quantum Alternating Operator Ansatz for  the Preparation and Detection \\ of Long-Lived Singlet States in NMR}

\author{Pratham Hullamballi$^\P$}
\email{pratham.hp@students.iiserpune.ac.in}

\author{Vishal Varma$^\P$}
\email{vishal.varma@students.iiserpune.ac.in}

\author{T. S. Mahesh}
\email{mahesh.ts@iiserpune.ac.in}

\affiliation{Department of Physics and NMR Research Center,\\
    Indian Institute of Science Education and Research, Pune 411008, India}

\keywords{QAOA, Long-lived state, Singlet state, NMR}

\begin{abstract}
Designing efficient and robust quantum control strategies is vital for developing quantum technologies. One recent strategy is the Quantum Alternating Operator Ansatz (QAOA) sequence that alternatively propagates under two noncommuting Hamiltonians, whose control parameters can be optimized to generate a gate or prepare a state. Here, we describe the design of a QAOA sequence to prepare long-lived singlet states (LLS) from the thermal state in NMR. With extraordinarily long lifetimes exceeding the spin-lattice relaxation time constant $T_1$, LLS have been of great interest for various applications, from spectroscopy to medical imaging. Accordingly, designing sequences for efficiently preparing LLS in a general spin system is crucial. Using numerical analysis, we study the efficiency and robustness of our QAOA sequence over a wide range of errors in the control parameters. Using a two-qubit NMR register, we conduct an experimental study to benchmark our QAOA sequence against other prominent methods of LLS preparation and observe superior performance, especially under noisy conditions. Finally, we numerically demonstrate the applicability of our QAOA sequence beyond two-qubit registers, specifically for polychromatic excitation of delocalized LLS in a six-proton system.
\end{abstract}

\def\thefootnote{$\P$}\footnotetext{These authors contributed equally to this work.}

\maketitle

\section{Introduction}

Variational quantum algorithms (VQAs) have emerged as promising candidates for the current noisy intermediate-scale quantum (NISQ) devices and have shown considerable advantage by making optimal use of both quantum resources and classical optimization techniques in what is now popularly known as a hybrid quantum-classical approach \cite{cerezo2021variational, bharti2022RMPnoisy, mcclean2016theory}. By transforming the optimization problem into a cost function measured on a quantum computer, a classical optimizer varies the parameters of a parametrized quantum circuit to minimize the cost.
A particular example of VQAs is the quantum approximate optimization algorithm, initially proposed by Farhi et al. \cite{farhi2014quantum} and later generalized to Quantum Alternating Operator Ansatz (QAOA) \cite{hadfield2019quantum}, which employs an alternating sequence of parametrized unitary transformations \cite{kostas2024reviewQAOA}.
Originally designed to solve combinatorial optimization problems such as MaxCut \cite{farhi2014quantum, zhou2020quantum,harrigan2021quantum, Farhi2022quantumapproximate}, QAOA has found numerous applications in preparing quantum many-body ground states of various Ising Hamiltonians \cite{wen2019Efficient,pagano2020pnasQuantum}. 
With their remarkable ability to prepare desired quantum states with shallow circuit depths and their utility for universal quantum control \cite{matos2021quantifying,lloyd2018quantum,morales2020universality}, QAOA has gained significant attention recently \cite{kostas2024reviewQAOA}. This paper investigates QAOA for quantum state preparation in nuclear magnetic resonance (NMR) quantum simulators.

In NMR spectroscopy, the spin-lattice relaxation time constant $T_1$ determines the rate at which a single spin attains thermal equilibrium from any nonequilibrium state \cite{book_cavanagh2007}. While an evolving nuclear magnetization captures crucial information about the surrounding physical environment, the $T_1$ process gradually restores its thermal equilibrium state, erasing all the information gathered during the dynamics.  Therefore, the $T_1$ timescale was long believed to be the rigid barrier beyond which no physical process may be studied using nuclear magnetization.  In a remarkable discovery two decades ago, Levitt and co-workers showed the preparation of the singlet order of a nuclear spin pair from thermal magnetization and demonstrated its extraordinarily long lifetime far beyond the $T_1$ barrier \cite{carravetta2004beyondT1, carravetta2004long}. 
Since then, the singlet order in a nearly symmetric spin pair has been popularly known as the Long-Lived State (LLS).  The long lifetime of the singlet state is a consequence of its immunity to intra-pair dipole-dipole relaxation, which forms the major source of relaxation in ordinary spin systems. However, it can not connect the antisymmetric singlet state to the symmetric triplet state \cite{book_pileio_2020}. 
In NMR, LLS has been extensively studied \cite{pileio2020long} and has found numerous applications such as chemical analysis \cite{cavadini2005slow}, biomedical imaging \cite{gloggler2017versatile, 2022_Sonnefeld_polyslic_sciadv}, protein-ligand binding \cite{salvi2012boosting, buratto2014exploring, buratto2016ligand}, and quantum information processing \cite{roy2010initialization}.  More recently, LLS has also been discovered in other architectures \cite{chen2017steady} and environments \cite{2013_ssNMR_lls, nagashima2014long, varma2023long}.

Over the years, several methods for LLS preparation have been developed, which include Carravetta-Levitt (hereafter CL) \cite{carravetta2004long}, M2S-S2M \cite{pileio2010storage}, SLIC \cite{devience2013preparation, 2022_Sonnefeld_polyslic_prl}, APSOC \cite{pravdivtsev2016robust,rodin2019constant}, symmetry-based sequences \cite{2022_Sabba_symmseq}, and optimal control \cite{wei2020preparation, deepak2017bang}.
The CL is a standard method for weakly/moderately coupled spins, whereas M2S-S2M, symmetry-based methods, and SLIC are suitable for strongly coupled spins \cite{pileio2020long, 2022_Sabba_symmseq}. The above-mentioned methods require precise delays and pulses and work only for weak or strongly coupled systems. On the other hand, adiabatic methods can work for both weak and strongly coupled systems and are robust against experimental imperfections \cite{rodin2019constant}. However, by nature, adiabatic methods require longer times, which may limit their efficiency.

\begin{figure}
\centering
\includegraphics[width=0.7\linewidth]{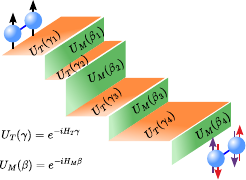}
\caption{Visualizing the Quantum Alternating Operator Ansatz (QAOA) method for quantum control tasks such as state preparation or gate synthesis. 
}
\label{fig:abstract}
\end{figure}

QAOA is a meta-heuristic algorithm based on a quantum gate-model that switches between unitaries selected from two types: phase-separation operators and mixing operators \cite{hadfield2019quantum}.
This work demonstrates QAOA as a general method for robust and efficient quantum state preparation by preparing LLS in a pair of nuclear spins (see Fig. \ref{fig:abstract}). We construct an experimentally feasible QAOA sequence and numerically analyze its performance for spins with different ranges of coupling strengths.
We then experimentally demonstrate and benchmark its performance against existing methods, such as APSOC, CL, M2S-S2M, and SLIC.

In Sec. \ref{sec:theory}, we first explain the theoretical framework of QAOA and subsequently set up the cost function for LLS preparation in terms of the parameters of the QAOA sequence.
In section \ref{Simulation}, we show the numerical simulation of LLS preparation using QAOA and analyze its feasibility and robustness 
in three types of systems: weakly/moderately coupled, strongly coupled, and very strongly coupled. In section \ref{Experiments}, we provide an experimental demonstration of QAOA in converting magnetization to singlet ($M\rightarrow S$) and singlet to magnetization ($S\rightarrow M$), and compare its performance against existing LLS preparation sequences. 
In section \ref{DelocalizedLLS}, we show prospects of using QAOA for chemically equivalent but magnetically inequivalent systems. 
Finally, we summarize and conclude in section \ref{Conclusion}.

\section{Theory}\label{sec:theory}
\subsection{QAOA}\label{ssec:qaoa}
As mentioned earlier, QAOA consists of an alternating sequence of two distinct operators. 
The phase-separation operators $U_T(\gamma) = e^{-i H_T \gamma }$ are parameterized by duration $\gamma$ and are generated by `target' Hamiltonian $H_T$, whose eigenstates encode the cost function.  The mixing operators $U_M(\beta) = e^{-i H_M \beta }$ are parameterized by duration $\beta$ and generated by a `mixer' Hamiltonian $H_M$, which does not commute with $H_T$. 
Therefore, starting from a convenient initial state $\ket{\psi_I}$, a QAOA circuit of $p$ layers creates a parameterized state with $2p$ parameters
\cite{farhi2014quantum, hadfield2019quantum}
\begin{align}
     \ket{\psi_F} = \ket{\psi (\vec{\gamma},\vec{\beta} )} &=  {U_{Q}}(\vec{\gamma},\vec{\beta}) \ket{\psi_I},~~\mbox{where,}
     \nonumber \\
      {U_{Q}}(\vec{\gamma},\vec{\beta}) &= \prod_{i=1}^{p} U_M(\beta_i) U_T(\gamma_i).
\end{align}
Using a classical optimizer, one can optimize the parameters $\vec{\gamma}$ and $\vec{\beta}$ to reach the ground state of $H_T$ by minimizing the cost function given by the energy  $\bra{\psi (\vec{\gamma},\vec{\beta} )} H_T \ket{\psi (\vec{\gamma},\vec{\beta} )}$.

As QAOA belongs to the larger family of variational quantum algorithms, it can also be used for 
state-to-state control\cite{matos2021quantifying}. 
In this case, there is more flexibility in choosing the set of alternating operators, 
and the QAOA unitary operator for $p$ layers will be
\begin{equation}
    {U_{Q}}(\vec{\gamma},\vec{\beta}) = \prod_{i=1}^{p} U_B(\beta_i) U_A(\gamma_i).
\end{equation}
Here $U_A(\gamma) = e^{-i H_A \gamma}$, $U_B(\beta) = e^{-i H_B \beta}$ are the unitaries generated by the two non-commuting Hamiltonian $H_A$ and $H_B$.
The cost function of the optimization problem can be redefined as the infidelity with the target state $\ket{\psi_T}$.
\begin{equation}
\mathcal{I}_T(\vec{\gamma},\vec{\beta}) = 1-\norm{\inpr{\psi_T}{\psi (\vec{\gamma},\vec{\beta})}}.    
\end{equation}

\subsection{Singlet Order Preparation}
Consider a pair of two interacting spin-1/2 particles.  Starting from a convenient initial state, the goal is to prepare the singlet order, which corresponds to the population difference between the singlet state $\ket{S_0} = (\ket{01}-\ket{10})/\sqrt{2}$ and the equally populated triplet states $\ket{T_0} = (\ket{01}+\ket{10})/\sqrt{2}$, $\ket{T_+} = \ket{00}$, and $\ket{T_-} = \ket{11}$.  
In practice, the degenerate triplet eigenstates of the isotropic Hamiltonian $-\mathbf{I}_1\cdot\mathbf{I}_2$ (with spin angular momentum operator $\mathbf{I}$) equilibrate rapidly to equalize their populations spontaneously \cite{carravetta2004long}.  
For an ensemble system such as an NMR register, if $\rho_I$ is the state after initialization, $\rho_{F} = {U_{Q}}(\vec{\gamma},\vec{\beta})\rho_I{U_{Q}}(\vec{\gamma},\vec{\beta})^\dagger$ is the final density matrix and $\rho_T$ is the target density matrix, the infidelity can be cast as \cite{fortunato2002strongly}
\begin{equation}
\mathcal{I}_{T}(\vec{\gamma},\vec{\beta}) = 1 - \frac{\tr(\rho_F \rho_T)}{\sqrt{\tr(\rho^2_T) Tr(\rho^2_F)}}.
\end{equation}
Under the high-temperature approximation in NMR, a density matrix can be written as $\rho = \mathbbm{1}/\tr[\mathbbm{1}] + \rho_\Delta$, where the first term represents the uniform background population that is invariant under the unitary dynamics, and the second term represents the trace-less deviation density matrix that captures all the interesting dynamics \cite{suter2008spins}.  Ignoring the identity term, the thermal state of a two-qubit NMR register is written as $I_{1}^z+I_{2}^z$, which after initialization by $U_I$ becomes $\rho_I = U_I (I_{1}^z+I_{2}^z) U_I^\dagger$, and the target state corresponding to the singlet order is $\rho_T = -\mathbf{I}_1\cdot\mathbf{I}_2$.

Using a $p$-layer QAOA ansatz, we can maximize the singlet content by minimizing the infidelity $\mathcal{I}_{S_0}(\vec{\gamma},\vec{\beta})$. In practice, it is also important to simultaneously minimize the total time $\sum_{i=1}^{p} (\gamma_i + \beta_i)$. Therefore, we proceed with linear scalarization and revise the cost function as
\begin{equation}\label{eq:cost_func}
    f({\vec{\gamma},\vec{\beta}}) = r \mathcal{I}_{S_0}(\vec{\gamma},\vec{\beta}) + (1-r) \sum_{i=1}^{p} (\gamma_i + \beta_i),
\end{equation}
where $r \in [0,1]$ is the positive real weight parameter.  For the singlet-order, the infidelity $\mathcal{I}_{S_0}$ has a lower bound of 
$1-\sqrt{2/3}$ \cite{soumya2010density, pileio2017methodology}.

\section{Numerical analysis of QAOA} \label{Simulation}
Consider a pair of spin qubits with internal NMR Hamiltonian (in a frame rotating at the average Larmor frequency)
\begin{equation}
    H_0 = -\pi \delta (I^z_1 - I^z_2) + 2\pi J \mathbf{I}_1 \cdot \mathbf{I}_2,
\end{equation}
where $\delta$ is the chemical shift difference, $J$ is the scalar coupling constant, and $I^z_i$ are the z-components of the spin angular momentum operators $\mathbf{I}_i$.

We choose the two alternating operators to be the evolution under the system Hamiltonian and the evolution with an RF field along the x-direction.
The explicit forms of the alternating Hamiltonians $H_A$ and $H_B$ for the QAOA sequence are
\begin{equation}{\label{eq:qaoa_Hamiltonian}}
    H_A = H_0  ~~\quad \textrm{and} \quad~~   H_B = H_0 - 2\pi\nu I^x -2\pi\Delta I^z.
\end{equation}
Here $I^\alpha = I^\alpha_1 + I^\alpha_2$ is the total $\alpha$-component of the spin operator. The initial state for the QAOA sequence is chosen to be $I_{1}^x+I_{2}^x$ since it allowed us to use lower-power RF pulses with $\Delta=0$ in QAOA layers.
In the Hamiltonian $H_B$, both the RF amplitude ($\nu$) and frequency offset ($\Delta$) are free parameters, which can be further optimized to gain better performance. 
The choice of these two particular Hamiltonians is motivated by the ease of experimental implementation, using delay as parameters, and lower RF power irradiation. But, in principle, one may use another separate set of alternating Hamiltonians.

Depending upon the ratio of $J$ to $\delta$, the coupling strength of the spin-pair can vary from weak to very strong \cite{book_cavanagh2007}.
In our numerical simulations, we consider three particular cases of $J$ and $\delta$ such that the resulting coupling strength is either moderate, strong, or very strong (see Tab. \ref{tab:systems}).
Tab. \ref{tab:systems} also lists the commonly preferred methods in each case to prepare and detect LLS. 
We numerically analyze the performance of our QAOA sequence with the 
Hamiltonians given in Eq. \ref{eq:qaoa_Hamiltonian} with different values of RF amplitudes and frequency offset, and for different cases of coupling strength.

\begin{table}[b]
\caption{\label{tab:systems}
Three types of spin systems based on the coupling strength to chemical shift ratios, their representative Hamiltonian parameters, and some preferred existing LLS preparation methods.  The parameters in the first row belong to the system chosen for experiments.}
\begin{ruledtabular}
\begin{tabular}{l cc l}
    $\begin{array}{c}\mbox{Coupling}\\\mbox{strength}\end{array}$         & \multicolumn{2}{c}{$\begin{array}{c}\mbox{Representative}\\\mbox{parameters}\end{array}$} & Existing methods \\
                        & $\delta$ (Hz) & $J$ (Hz)  \\  \colrule
    Moderate      &   35.8        & 17.2 & CL, M2S, APSOC \\
    Strong  &   10.0        & 18.0 & M2S, APSOC, SLIC \\
    Very strong    &   10.0        & 54.0 & M2S, APSOC, SLIC  \\        
    \end{tabular}
\end{ruledtabular}
\end{table}

\begin{figure}
\centering
\includegraphics[width=0.95\linewidth]{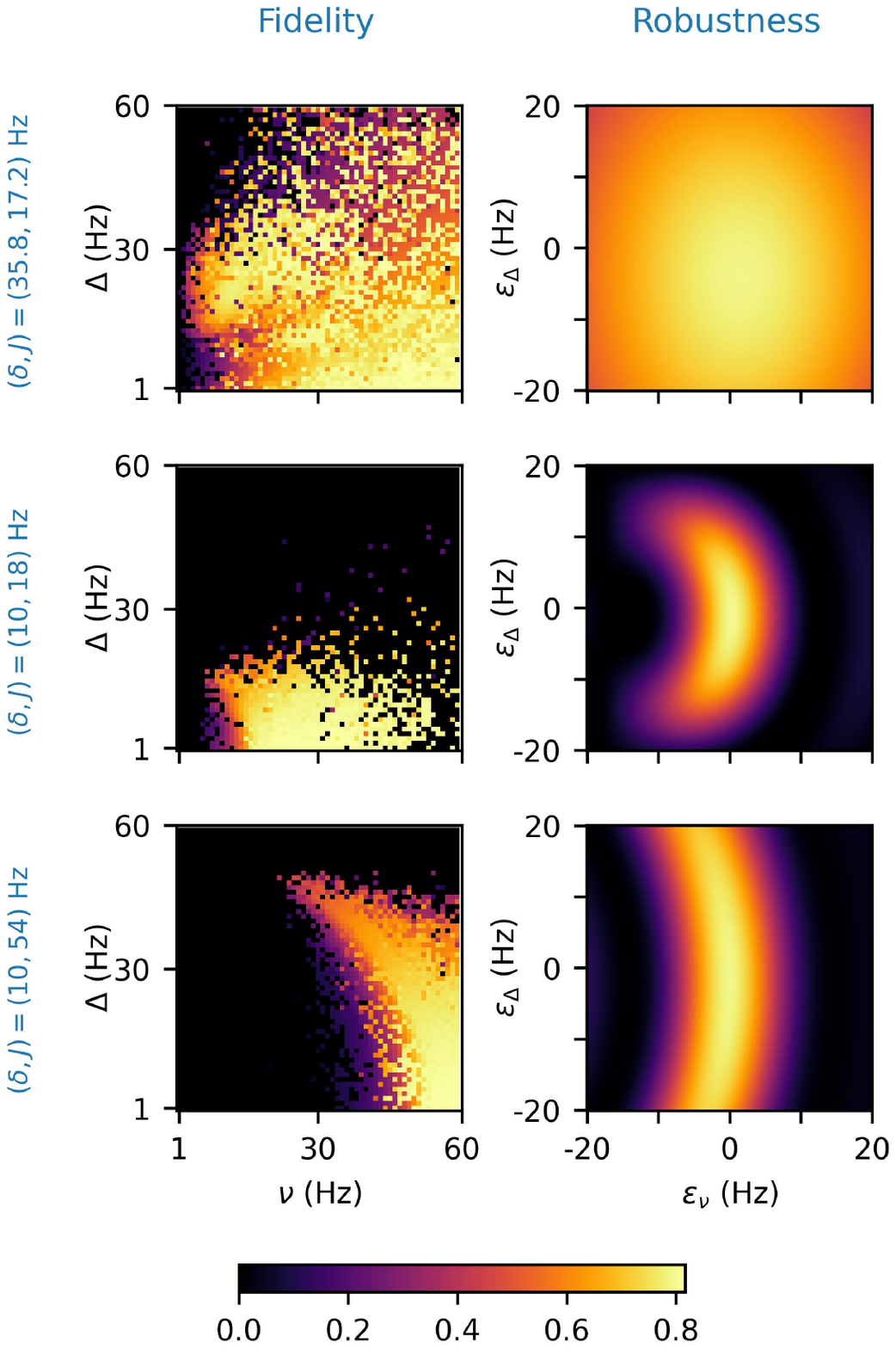}
\caption{LLS preparation fidelity heatmaps for four-layered QAOA sequence obtained by finding the optimal parameters for given $\nu$ and $\Delta$ values. QAOA robustness is plotted against deviations $\epsilon_{\Delta}$ and $\epsilon_{\nu}$ (from optimal $\Delta$ and $\nu$ values) introduced during preparation. 
The optimal ($\nu$, $\Delta$) about which deviations are considered are 
$(58, 4)$ Hz, $(18, 1)$ Hz, and $(19, 1)$ Hz 
for moderate, strong, and very strong coupling cases, respectively. We have used $r=0.4$ (see Eq. \ref{eq:cost_func}) for finding optimal parameters.
}
\label{fig:simulation}
\end{figure}

The fidelity heatmaps in Fig. \ref{fig:simulation} show the feasibility ranges of control parameters $\nu$ and $\Delta$ for LLS preparation by QAOA.  They indicate that LLS preparation is achievable in a wider range of control parameters in a moderately coupled system compared to a strongly or very strongly coupled system. 
We see that for the moderately coupled system ($\delta=35.8$ and $J=17.2$), the numerical optimization mostly results in successfully finding a set of optimal $\vec{\gamma}$ and $\vec{\beta}$ 
parameters for most values of RF amplitude and frequency offset (depicted by the bright region of the heatmap in Fig. \ref{fig:simulation}). In addition to the conversion efficiency, we observe that the QAOA is very robust against RF amplitude and offset errors.

For the strongly coupled case ($\delta=10$ and $J=18$), we see that for a certain range of $\Delta$ and $\nu$, the numerical optimizer is able to find the optimal 
$\vec{\gamma}$ and $\vec{\beta}$
parameters. For the very strongly coupled system ($\delta=10$ and $J=54$), the range of values for $\nu$ and $\Delta$ shifts towards higher RF amplitudes.
Even in strongly and very strongly coupled cases, QAOA remains resilient to errors in RF frequency offset $\Delta$, while less so for errors in RF amplitude $\nu$.

\begin{figure}[t]
\centering
\includegraphics[width=\linewidth]{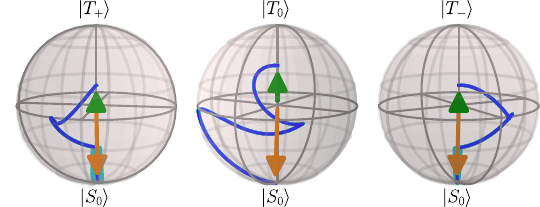}
\caption{The Bloch sphere trajectory (in $\{\ket{S_0},\ket{T_{0/\pm}}\}$ basis) for our QAOA sequence with parameters in Tab. \ref{tab:qaoa_expt}.  
The dark line corresponds to the nominal control amplitude, and the associated patch corresponds to $\pm 20 \%$ deviated control-field amplitude (RF inhomogeneity).
In each sphere, the up-arrow corresponds to the initial pure $\ket{++}$ state, and the down-arrow indicates the final state after the QAOA sequence. }
\label{fig:bloch_sphere}
\end{figure}

To gain an insight into the dynamics under the QAOA sequence, Fig. \ref{fig:bloch_sphere} plots magnetization trajectories in the $\{\ket{S_0},\ket{T_{0/\pm}}\}$ Bloch-spheres. 
The resilience to error in RF amplitude can also be seen in Fig. \ref{fig:bloch_sphere}. The trajectories that have a non-zero error in RF amplitude do not diverge from the path taken by the error-free trajectory. Near the end of all trajectories (close to the $\ket{S_0}$), we see a very small deviation from the actual intended path.

\begin{table}[b]
\caption{\label{tab:qaoa_expt}
QAOA Hamiltonians and optimized parameters for magnetization-to-singlet ($M\rightarrow S$) conversion and singlet-to-magnetization ($S\rightarrow M$) reconversion in a moderately coupled system ($\delta=35.8$ Hz, $J=17.2$ Hz).  
We have used Constrained Optimization By Linear Approximation (COBYLA) implemented in Python's \textit{SciPy} \cite{2020SciPy}. 
The parameters for the Hamiltonian in Eq. \ref{eq:qaoa_Hamiltonian} are taken as $\nu = \nu' = 100$ Hz and $\Delta = \Delta' = 0$.
}
\begin{ruledtabular}
    \begin{tabular}{c cc cc}
        \multicolumn{1}{c}{\textrm{Layer}} & \multicolumn{2}{c}{\textrm{$M\rightarrow S$}}
        & \multicolumn{2}{c}{\textrm{$S\rightarrow M$}}
        \vspace{0.1cm}  \\ \cline{2-3} \cline{4-5} \vspace{0.1cm}
          & $\gamma_i$ & $\beta_i$ &  $\gamma_i'$ & $\beta_i'$   \\
          &  (ms)     &  (ms)      &   (ms)      &    (ms)       \\
        \colrule
        1 & 14.069 & 3.452 & 2.457 & 6.255  \\
        2 &  6.810 & 0.025 & 6.401 & 2.420  \\
    \end{tabular}
\end{ruledtabular}
\end{table}

A similar method called homonuclear-ADAPT \cite{elliott2019homonuclear} also uses an alternating sequence of fixed-angle hard pulses and delays to prepare LLS, where the number of repetitions and duration of delay can be obtained by fixing the angle of the hard pulse in the case of the strongly coupled system. Though the homonuclear-ADAPT method does not require numerical optimization for the pulse parameters in the case of a strongly coupled system, it does need numerical optimization for weakly coupled cases, as the assumptions taken to derive the formula are not applicable in weak coupling cases. Homonuclear-ADAPT also requires high power pulses and a large number of repetitions, because of which errors may accumulate and the LLS-preparation efficiency may be reduced. QAOA, on the other hand, uses low-power RF and less number of repetitions.

\section{Experiments}\label{Experiments}

\begin{figure}
    \centering
    \includegraphics[width=0.85\linewidth]{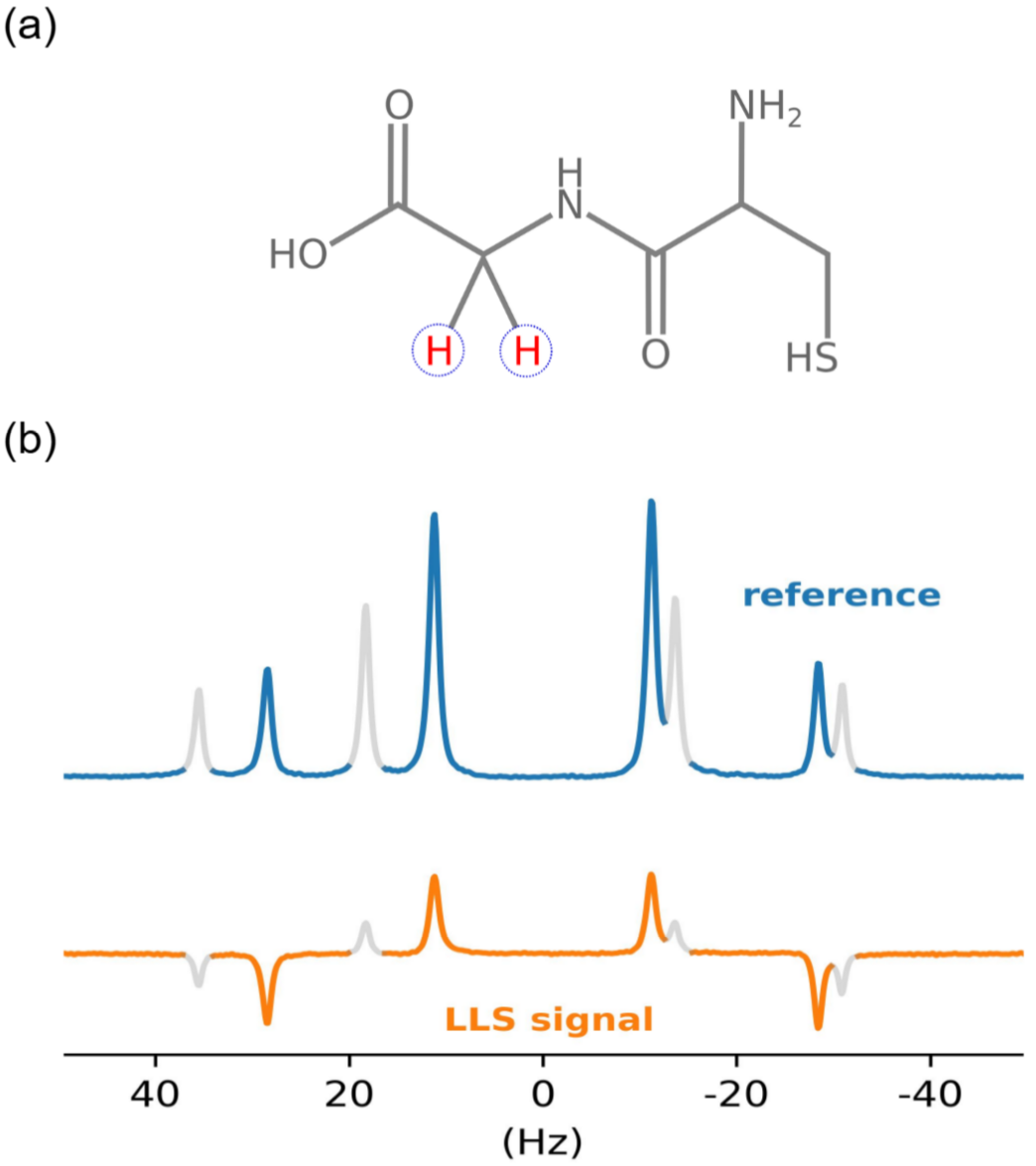}
    \caption{(a). The spin system used in this work. The two proton spins used are shown in circles. (b) The reference NMR spectrum and the LLS spectrum obtained using the QAOA sequence
    (with 15 seconds of storage time) are shown. 
    The dimmed peaks are presumably from another unrelated AB system.
    }
    \label{fig:mol_spec}
\end{figure}

\begin{figure}
\centering
\includegraphics[width=\linewidth]{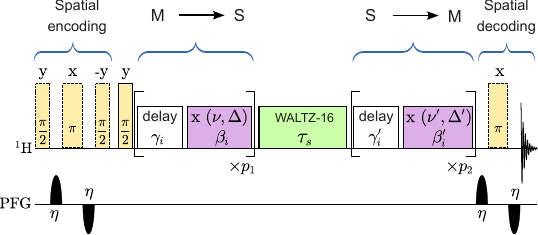}
\caption{The QAOA  sequence for magnetization to singlet-order transfer (M$\rightarrow $S) and singlet-order to magnetization transfer (S$\rightarrow $M). 
Here, $\gamma_i^{(}$$'$$^{)}$ and $\beta_i^{(}$$'$$^{)}$ are respectively durations of target and mixer elements
The parameters $\nu^{(}$$'$$^{)}$ and $\Delta^{(}$$'$$^{)}$ are RF pulse strength and RF offset, as described in Table \ref{tab:qaoa_expt}. 
The $(\pi/2)_y$ pulse before 
M$\rightarrow$S is for initialization.
The spatial encoding and spatial decoding parts are used to introduce noise by means of the diffusion of nuclear spins. $\eta$ is the gradient duration. 
The final observable state 
that we have chosen
is the single quantum magnetization $I_1^x I_2^z - I_1^z I_2^x$ that generates a characteristic anti-phase NMR signal \cite{carravetta2004long}.
In principle, one can choose a different final state.
}
\label{fig:nmrseq}
\end{figure}
We now benchmark the performance of the QAOA sequence against standard LLS sequences, 
namely CL \cite{carravetta2004long}, M2S-S2M \cite{pileio2010storage}, APSOC \cite{rodin2019constant}, and SLIC \cite{devience2013preparation}.
All experiments were done on a 9.4 Tesla Bruker Avance-III NMR spectrometer at an ambient temperature of 298 K.
Our system comprises proton spin pairs of Cys-Gly dipeptide (see Fig. \ref{fig:mol_spec}a). The sample was prepared by dissolving 4mg of Cys-Gly in 700 $\mu$L D$_2$O and bubbled with Argon gas to remove dissolved oxygen.
The two nuclear spins are coupled to each other with $J = 17.2$ Hz and with the chemical shift difference $\delta = 35.8$ Hz (see first row of Tab. \ref{tab:systems}).
We obtained the longitudinal relaxation time constant $T_1 = 1.6$ s from the inversion recovery experiments for both spins.  In all experiments, we have used a 1 kHz WALTZ-16 spin-lock sequence to sustain LLS during storage. 
The optimized experimental parameters for QAOA are given in Tab. \ref{tab:qaoa_expt}.
The spectrum corresponding to LLS obtained using QAOA and the thermal (reference) spectrum are shown in Fig. \ref{fig:mol_spec}b.

The standard LLS preparation methods described above are designed with certain assumptions about the strength of the coupling between the nuclear spins. For example, the CL method works well in the weak to moderate coupling range, whereas M2S-S2M and SLIC are more suitable for strongly coupled systems. Therefore, to make a fair comparison, we optimize the parameters for each of these methods to obtain the highest possible preparation of LLS with minimum time. The optimal parameters used in experiments for each LLS method are given in Ref. \cite{supp}.

For all the methods mentioned above, we test their performance in transferring magnetization-to-singlet and singlet-to-magnetization under different noise levels. To introduce noise, we use diffusion of the spin systems and vary the strength of the pulse-field gradients (PFG) used for spatial encoding and decoding of the spin systems. We have considered three cases for PFG strength: low noise (0\%), medium noise (23\%), and strong noise (47\%) \cite{varma2023long, book_cavanagh2007}. The pulse sequence used to introduce noise is shown in Fig. \ref{fig:nmrseq}.
We have used a sinusoidal pulse-field gradient of 200 $\mu$s duration in all the experiments where the noise is introduced.

\begin{table}
\caption{\label{tab:pulse_time}
LLS preparation ($M\rightarrow S$) and detection ($S\rightarrow M$) time for the weakly coupled system ($\delta=35.8$ Hz, $J=17.2$ Hz) taken by all methods considered in this work.
}
\begin{ruledtabular}
    \begin{tabular}{l c c c}
        Method  & \multicolumn{3}{c}{Time (ms)} \\ \cline{2-4}
                & $M \rightarrow S$ & $S\rightarrow M$ & Total        \\ \colrule
        QAOA    & 24.366            & 17.503           &  41.869      \\
        APSOC   & 160.000           & 160.010          & 320.010      \\
        CL      & 133.040           & 6.310            & 139.350      \\
        SLIC    & 21.510            & 21.500           &  43.010      \\
        M2S-S2M & 63.005            & 62.995           & 126.000
    \end{tabular}
\end{ruledtabular}
\label{tab.durations}
\end{table}

\begin{figure*}[t]
\centering
\includegraphics[width=0.95\linewidth]{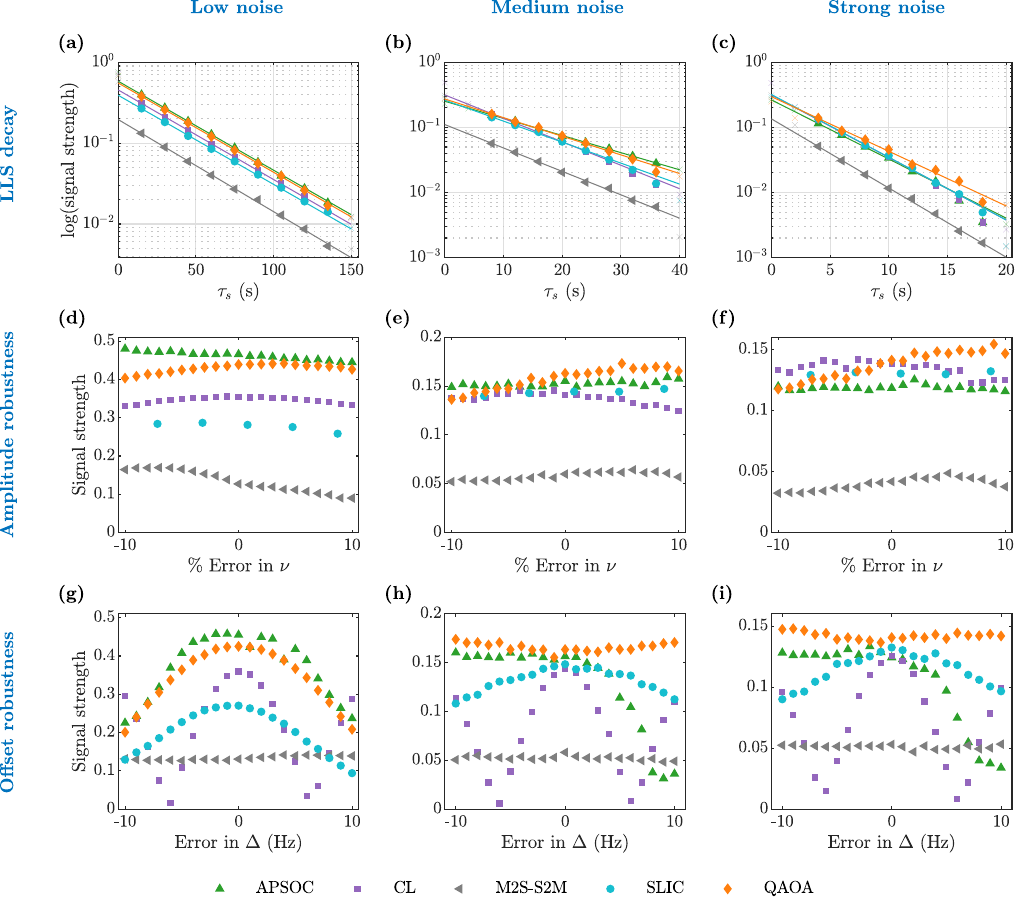}
\caption{Experimental results for benchmarking the QAOA sequence against the standard LLS preparation sequences for three different levels of noise cases. 
Decay of LLS with varying storage time in case of low noise (a), medium noise (b), and strong noise (c). 
The LLS life-times $T_{LLS}$ averaged over different methods obtained for three different noise ranges are 39.1 $\pm$ 0.7 s, 13.9 $\pm$ 1.0 s, and 4.6 $\pm$ 0.4 s (see Ref.\cite{supp}).
Plots (d)-(f) show the robustness of all methods against deviation in the RF amplitude for low noise (d), medium noise (e), and strong noise (f). Plots (g)-(i) show the RF offset robustness of all methods for low noise (g), medium noise (h), and strong noise (i). For all robustness experiments, the LLS storage times are: 10 seconds for the low noise case, 8 seconds for the medium noise case, and 4 seconds for the strong noise case.
}
\label{fig:experiments_results_all}
\end{figure*}

To benchmark the conversion efficiency of our QAOA method against other methods, we plot the decay of the singlet-order with the varying LLS storage time. For the low noise case (i.e., no PFG is used), we see that QAOA is performing much better than M2S-S2M, SLIC, and CL. QAOA and APSOC give similar performances in terms of the final fidelity of the singlet order (see Fig. \ref{fig:experiments_results_all}a); however, we note that the QAOA sequence takes much less duration compared to the APSOC sequence (see Tab. \ref{tab.durations}).
Fig. \ref{fig:experiments_results_all}b and \ref{fig:experiments_results_all}c compare the decay of singlet order in the case of medium and strong noise, respectively, for all LLS methods. We see that QAOA still performs better or is comparable to most methods.

Next, we test the robustness of the methods against errors in RF amplitude and frequency offset under the three noise cases. 
For testing the robustness against RF amplitude, we have varied the amplitude of the RF pulse from $-10\%$ to $+10\%$ of the optimal RF amplitude in each LLS method. For the case of robustness against RF offset, we introduce an error of $-10$ Hz to $+10$ Hz in the optimal value of RF for each LLS method. 
Fig. \ref{fig:experiments_results_all}d compares the robustness of all methods against the amplitude errors in the low noise case. We observe that APSOC and QAOA perform better than all other sequences up to $\pm10\%$ error in RF amplitude.
In the case of medium noise, we see that QAOA performs better than all other sequences overall (see Fig. \ref{fig:experiments_results_all}e).
For the strong noise case, QAOA still performs better overall than most other sequences (see Fig. \ref{fig:experiments_results_all}f).

For the case of offset robustness, APSOC and QAOA show the best performance among all methods in the case of low noise (Fig. \ref{fig:experiments_results_all}g). For the medium and strong noise, we see that QAOA is the most robust among all other sequences (see Fig. \ref{fig:experiments_results_all}h and \ref{fig:experiments_results_all}i) up to a deviation of $\pm10$ Hz in the RF offset.

These results clearly establish the superior utility of the QAOA sequence with respect to the efficiency of LLS preparation under different noisy conditions, as well as in terms of the pulse duration.
Once again, QAOA shows the best performance against offset deviations by as much as $\pm 10$ Hz, and it is also best overall in amplitude deviations by as much as $\pm 10\%$, although it is somewhat sensitive to lesser pulse amplitude.
We notice a general agreement between the experimental robustness plots of Fig. \ref{fig:experiments_results_all} with the simulation robustness plots of Fig. \ref{fig:simulation}. 

\section{QAOA for chemically equivalent but magnetically inequivalent spin pairs}\label{DelocalizedLLS}
\begin{figure}
    \centering
    \includegraphics[width=0.98\linewidth]{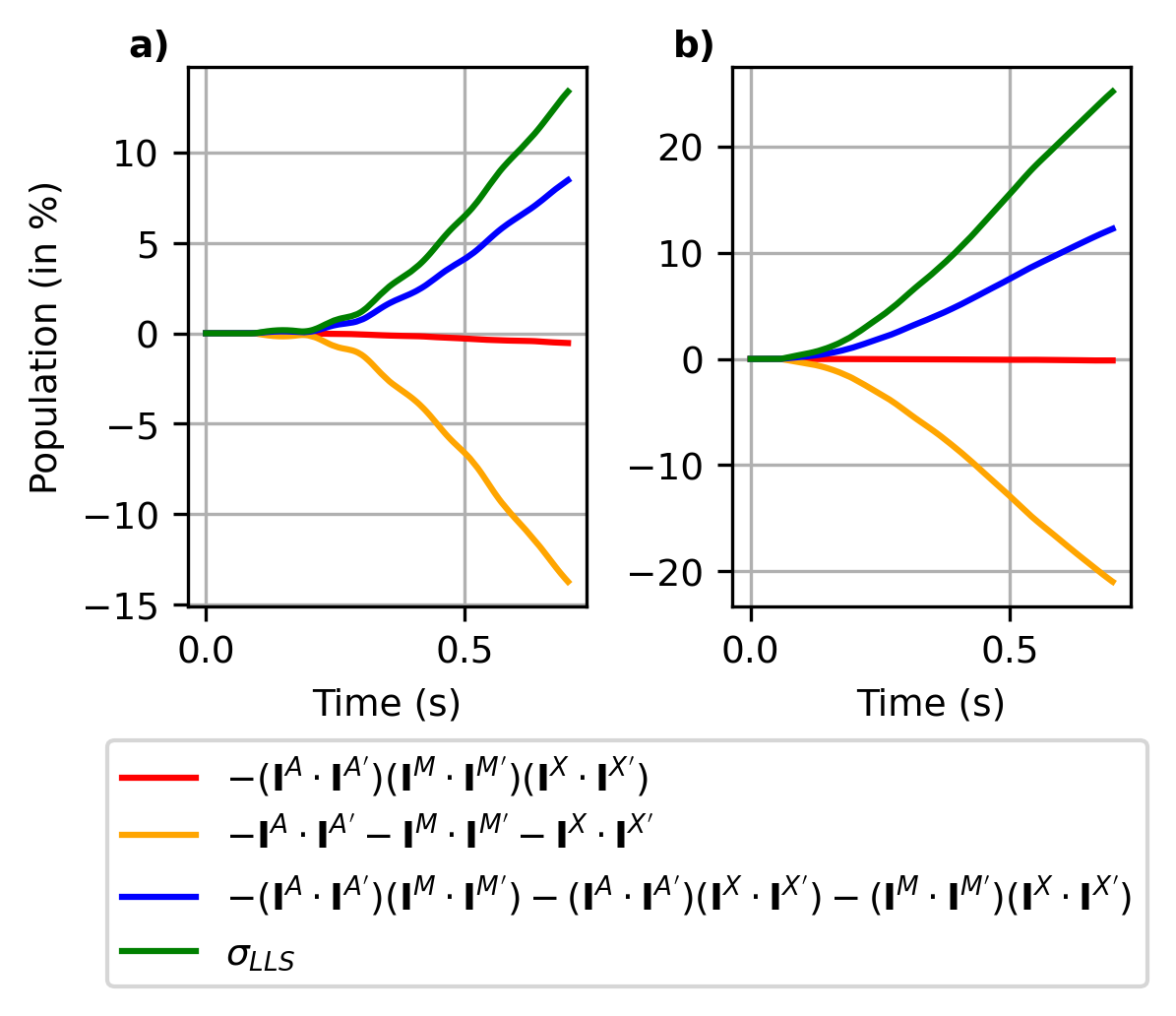}
    \caption{Numerical simulations of the excitation of various components in the target state $\sigma_{LLS}$ associated with (a). ``DQ triple SLIC" method and (b). ``SQ double SLIC" method \cite{2022_Sonnefeld_polyslic_prl}. For details of the simulations, see Ref. \cite{supp}.}
    \label{fig:polyLLS_QAOA}
\end{figure}

QAOA may also be used for 
preparing LLS in systems with chemical equivalence but magnetic inequivalence, such as $AA'XX'$, $AA'MM'XX'$, and so on \cite{feng2012accessing, 2022_Sonnefeld_polyslic_prl, 2022_Sonnefeld_polyslic_sciadv}.
In the following, we show numerical results for QAOA on the $AA'MM'XX'$ system with the same parameters described by Sonnefeld et al. \cite{2022_Sonnefeld_polyslic_prl}. 
For the two alternating Hamiltonian operators, we have used the natural Hamiltonian of the $AA'MM'XX'$ as $H_A$ and 
    $H_B = H_A - 2\pi \nu I_x  - 2\pi\Delta I_z$,
with global transverse and longitudinal operators $I_x$ and $I_z$ respectively.
The initial and final states are set to $\rho_{I} = I_z$ and $\rho_T = \sigma_{LLS}$ respectively, where the delocalized LLS density matrix $\sigma_{LLS}$ is described in Ref. \cite{2022_Sonnefeld_polyslic_prl}.
Using a 4-layer QAOA with a total duration of 0.7 s \cite{supp}, we could achieve the polychromatic excitation of delocalized LLS in the $AA'MM'XX'$ system. The evolutions of populations of various components during QAOA are plotted in FIG.~\ref{fig:polyLLS_QAOA}. 

\section{Discussions and Conclusions} \label{Conclusion}
Robust quantum control is crucial for future quantum technologies, and accordingly, numerous quantum control methods have been developed \cite{mahesh2023quantum}.  The recent QAOA method comprises alternating unitaries generated by two Hamiltonians whose parameters are classically optimized for efficient gate or state synthesis.  In some sense, QAOA generalizes over the bang-bang quantum control, which has been demonstrated earlier \cite{gaurav2016steering, yang2017pontryagin, liang2020investigating}.
Here, we have demonstrated the superior performance of QAOA for state preparation, specifically in preparing the long-lived singlet state (LLS).  Since its discovery, LLS has found several applications, from spectroscopy to medical imaging to quantum information.

We have designed a QAOA sequence and experimentally compared its performance against other standard LLS preparation methods, such as CL, APSOC, M2S-S2M, APSOC, and SLIC. 
We experimentally analyzed the performance of these sequences under the effect of noise with varying strength. We found that in noisy cases, QAOA performs better or is comparable to other methods of LLS preparation. In addition to being very efficient, QAOA is also much shorter compared to the other LLS sequences described here. Furthermore, the standard methods of LLS preparations work only in specific coupling regimes. For example, CL works well for weakly coupled cases, whereas M2S-S2M and SLIC work well in the case of strongly coupled cases. QAOA, on the other hand, can work for a broad range of coupling strengths. QAOA offers more flexibility in choosing experimentally feasible operators with control over the parameters of the mixer and target Hamiltonian. Another advantage of QAOA is the use of low-power pulses for preparing LLS, which also alleviates the problem of heating due to strong RF pulses. 
We have made extensive numerical analyses to study the feasibility and robustness of the QAOA sequence for different ranges of system parameters and control field parameters. 
Our numerical studies confirm that QAOA can efficiently prepare LLS across all systems, whether weakly or strongly coupled. Additionally, our numerical simulations demonstrate that 
QAOA can also be extended to systems with chemical equivalence but magnetic inequivalence, such as $AA'XX'$ and $AA'MM'XX'$, 
highlighting its applicability beyond simple two-qubit cases.

We hope such efficient sequences for conversion and reconversion of magnetization to LLS will advance the scope of LLS applications.
We also envisage the applications of QAOA as a general quantum control protocol for various tasks in quantum computing and other fields, such as spectroscopy, imaging, etc.

\section{Supplementary Material}
The supplementary material contains the details of the optimized parameters of the LLS sequences that are used to benchmark the QAOA sequence. It also contains the simulation results of QAOA for the conversion of LLS to magnetization. 

\section*{Acknowledgments}
We are grateful to Mr. Nitin Dalvi of IISER Pune for helping with sample degassing and to Mr. Pranav Chandarana of the University of the Basque Country for valuable discussions.
The DST/ICPS/QuST/2019/Q67 funding is gratefully acknowledged. We also thank the National Mission on Interdisciplinary Cyber-Physical Systems for funding from the DST, Government of India, through the I-HUB Quantum Technology Foundation, IISER-Pune.

\section*{Author Declarations}

\subsection*{Conflict of Interest}
The authors have no conflicts to disclose. 

\subsection*{Author Contributions}
\textbf{Pratham Hullamballi}: Data curation (equal); Formal analysis (equal); Methodology (equal); Validation (equal); Visualization (equal); Writing – original draft (equal); Writing – review \& editing (equal). \textbf{Vishal Varma}: Formal analysis (equal); Investigation (equal); Methodology (equal); Validation (equal); Visualization (equal); Writing – original draft (equal); Writing – review \& editing (equal). \textbf{T. S. Mahesh}: Conceptualization (equal); Funding acquisition (equal); Resources (equal); Supervision (equal); Writing – review \& editing (equal).

\section*{Data Availability}
The data that support the findings of this study are available from the corresponding author upon reasonable request.

\bibliography{ref}


\pagebreak
\widetext
\begin{center}
\textbf{\large Supplemental Material}
\end{center}
\setcounter{equation}{0}
\setcounter{figure}{0}
\setcounter{table}{0}
\setcounter{page}{1}
\setcounter{section}{0}

\makeatletter
\renewcommand{\theequation}{S\arabic{equation}}
\renewcommand{\thefigure}{S\arabic{figure}}
\renewcommand{\bibnumfmt}[1]{[S#1]}
\renewcommand{\citenumfont}[1]{S#1}
\renewcommand{\thesection}{S-\Roman{section}}
\renewcommand{\thetable}{S\arabic{table}}

\section{Optimal Parameters for CL}
\noindent
We find the optimal delay parameters for the CL sequence by using brute-force search through the discretized parameter set of delays with a step increment of 1 millisecond. We take the parameters that prepare LLS in minimum duration with final state fidelity above a certain threshold (in this case, 0.99). The optimal parameters used for the CL sequence in this work are given below.\\
$\tau_1 = 0.043 $ s \\
$\tau_2 = 0.083 $ s \\
$\tau_3 = 0.007 $ s \\
$\tau_5 = 0.0063 $ s \\
The total time taken by the CL sequence is then 0.139350 s.

\section{Optimal Parameters for M2S-S2M}
\noindent
The optimal $n_1$ and $n_2$ number of repetitions of the elements ($\tau_d - \pi - \tau_d$) and the optimal delay time $\tau_d$ in the M2S-S2M sequence are found by using a brute force search. The optimal parameters used in the experiments are given below.
$\tau_d = 0.012589 $ s \\
$n_1 = 1 $ \\
$n_2 = 1 $ \\
Total pulse time (preparation + detection) : 0.063005 + 0.062995 = 0.126000 s.

\section{Optimal Parameters for APSOC}
\noindent
To find the optimal APSOC pulse, we optimize over the constant RF offset, which must be non-zero. We choose the one that produces the highest fidelity in minimum time for all possible RF offsets. The optimal parameters used in the experiments are:
The offset, $\Delta = 20 $ Hz. \\
Pulse duration $\tau = 0.16 $ s. \\
Max pulse amplitude = 281 Hz. \\
Total time for LLS preparation + detection = 0.160000 + 0.160010 = 0.320010 s.

\section{Optimal Parameters for SLIC}
\noindent
For SLIC also, we follow the same approach as before and perform brute-force optimization over the RF strength of the low-power spin lock pulse and its duration. The optimal duration and the strength for the spin-lock in used in the SLIC experiments are:\\
Pulse duration $\tau_p = 21.5 $ ms. \\
Pulse amplitude $\nu = 25.3 $ Hz.

\section{Fidelity Heatmap and Robustness for Detection Phase and Overall Protocol}
\noindent

To further demonstrate QAOA as a useful detection protocol (S$\rightarrow$M), fidelity heatmaps and robustness plots for converting singlets to single quantum magnetisation are plotted in Fig. S1.

\begin{figure*}[h]
\centering
\includegraphics[width=\linewidth]{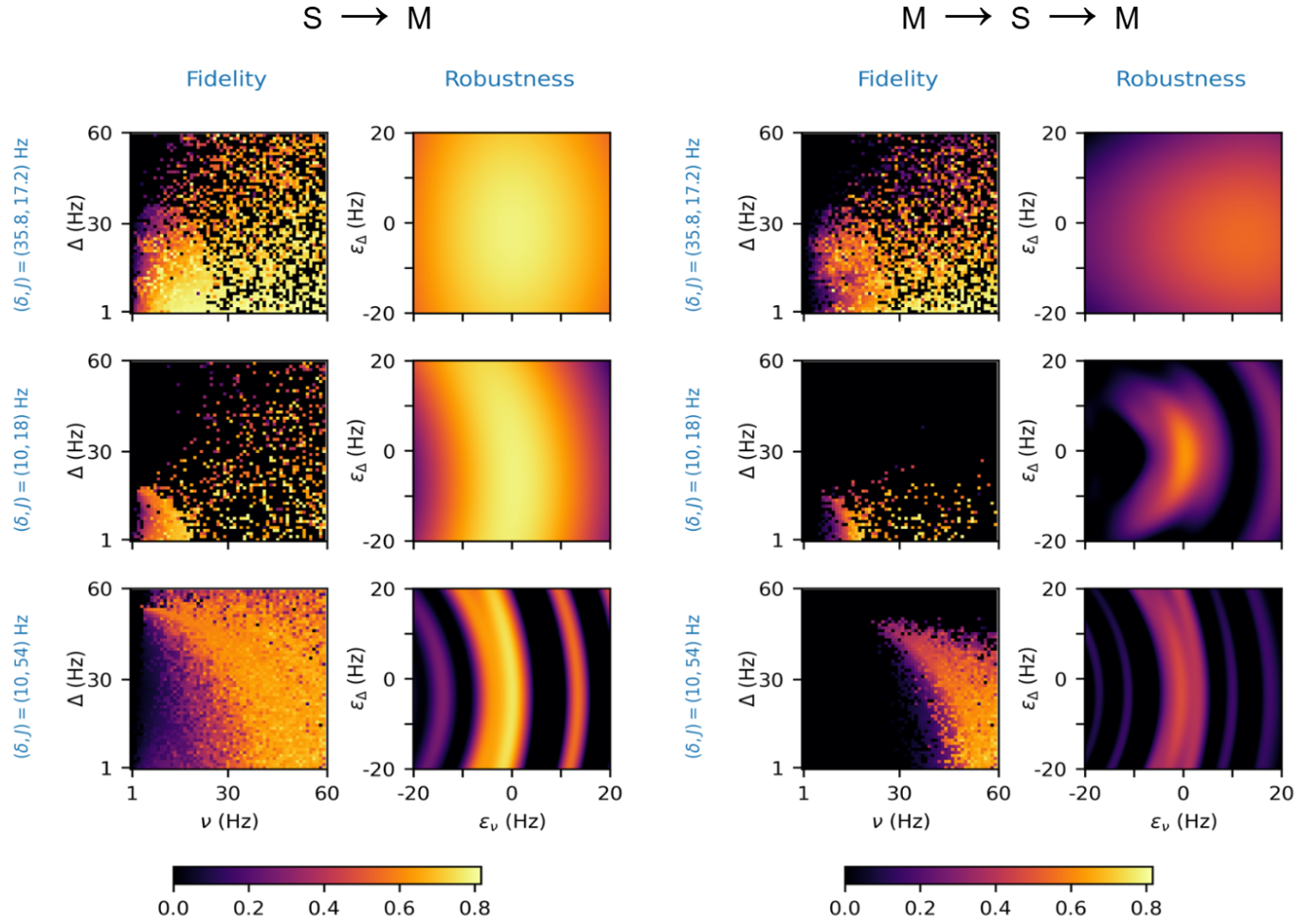}
\caption{Fidelity heatmap for 4-layered QAOA sequence obtained by finding the best solution for given $\nu$ and $\Delta$ values for the detection phase and total protocol. QAOA robustness against deviations $\epsilon_{\Delta}$ and $\epsilon_{\nu}$ (from optimal $\Delta$ and $\nu$ values) introduced during $S \rightarrow M$ and $M \rightarrow S \rightarrow M$ protocol. For $S \rightarrow M$ case, the optimal ($\nu$, $\Delta$) about which deviations are considered are $(42~\text{Hz}, 2~\text{Hz})$, $(59~\text{Hz}, 7~\text{Hz})$ and $(52~\text{Hz}, 10~\text{Hz})$ for moderate, strong and very strong coupling cases.}
\label{fig:detection_total}
\end{figure*}

Keeping both $\nu$ and $\Delta$ same for both preparation (M$\rightarrow$S) and detection (S$\rightarrow$M), we plot the overall fidelity for the whole sequence (M$\rightarrow$S$\rightarrow$M) in Fig. S1. The robustness plots for the final remaining signal are produced by considering the same magnitude of error in $\Delta$ and $\nu$ in both preparation and detection protocols. 
Fidelity heatmaps in $M \rightarrow S \rightarrow M$ assume RF parameters to be fixed in both preparation and detection. However, for a better performance one can choose different optimal RF parameters for the preparation part and detection part.

\section{The LLS decay constant, $T_{LLS}$, obtained by different methods}
\noindent
To calculate the lifetime of the singlet state, we fit an exponential function to the signal amplitude obtained by varying the storage time (shown in Fig. 6 of the main text).

\begin{table}[h!]
\caption{\label{tab:tlls_values}
$T_{LLS}$ values obtained by fitting a decaying exponential to the final NMR signal amplitude using different methods by varying the storage time.
}
\begin{ruledtabular}
\begin{tabular}{l ccc}
    Method  & Low noise      & Medium noise   & Strong noise   \vspace{0.1cm}\\
    \colrule \\
    APSOC   & 39.6 $\pm$ 0.4 & 16.6 $\pm$ 0.7 &  4.8 $\pm$ 0.3 \\
    CL      & 39.2 $\pm$ 0.4 & 12.1 $\pm$ 0.8 &  4.5 $\pm$ 0.2 \\
    SLIC    & 38.1 $\pm$ 0.7 & 12.1 $\pm$ 0.5 &  4.0 $\pm$ 0.1 \\
    M2S-S2M & 39.4 $\pm$ 0.7 & 13.5 $\pm$ 1.1 &  4.5 $\pm$ 0.2 \\
    QAOA    & 39.4 $\pm$ 0.3 & 15.2 $\pm$ 1.1 &  5.2 $\pm$ 0.4 \\
    \end{tabular}
\end{ruledtabular}
\end{table}

\newpage
\section{Optimal Parameters for Polychromatic Excitation of Delocalized LLS in Aliphatic Chains using QAOA}

\begin{table}[h]
    \centering
    \caption{\textbf{DQ triple SLIC case}: Parameters for the Hamiltonian $H_B$ are $\nu = 1833.71$ Hz and $\Delta = 0$ Hz. The achieved fidelity is 0.1338, with a total preparation time of 0.699 seconds.}
    \begin{tabular}{@{}ccc@{}}
        \toprule
        \textbf{Layer} & $\boldsymbol{\gamma_i}$ (ms) & $\boldsymbol{\beta_i}$ (ms) \\
        \midrule
        1 & 0.325   & 22.002  \\
        2 & 64.867  & 11.286  \\
        3 & 83.491  & 517.7018 \\
        \bottomrule
    \end{tabular}
    \label{tab:dq_slic}
\end{table}

\begin{table}[h!]
    \centering
    \caption{\textbf{SQ double SLIC case}: Parameters for the Hamiltonian $H_B$ are $\nu = 1933.02$ Hz and $\Delta = 0$ Hz. The achieved fidelity is 0.252, with a total preparation time of 0.699 seconds.}
    \begin{tabular}{@{}ccc@{}}
        \toprule
        \textbf{Layer} & $\boldsymbol{\gamma_i}$ (ms) & $\boldsymbol{\beta_i}$ (ms) \\
        \midrule
        1 & 0.663  & 32.936  \\
        2 & 24.021 & 0.8290  \\
        3 & 8.212  & 630.515 \\
        4 & 0.907  & 1.252   \\
        \bottomrule
    \end{tabular}
    \label{tab:sq_double_slic}
\end{table}

\end{document}